\input phyzzx
\hoffset=0.375in
\overfullrule=0pt

\def\etal{et al.\ }

\def\mpc{{\rm Mpc}}
\def\dol{{D_{\rm OL}}}
\def\dls{{D_{\rm LS}}}
\def\dos{{D_{\rm OS}}}

\def\kms{\, {\rm km}\,{\rm s}^{-1}}
\def\muas{\, \mu {\rm as}}
\def\msun{\, M_{\odot}}

\def\bv{{\bf v}}
\def\bvt{{\bf v}_t}
\def\bvs{{\bf v}_s}
\def\bvo{{\bf v}_o}
\def\bvl{{\bf v}_l}

\twelvepoint
\font\bigfont=cmr17
\centerline{\bigfont Signatures of Accretion Disks in Quasar Microlensing}
\bigskip
\centerline{{\bf Andrew Gould}\footnote{1}{Alfred P.\ Sloan Foundation Fellow}}
\smallskip
\centerline{Dept of Astronomy, Ohio State University, Columbus, OH 43210}
\smallskip
\centerline{{\bf Jordi Miralda-Escud\'e}}
\smallskip
\centerline{Dept of Physics \& Astronomy, University of Pennsylvania, 
Philadelphia, PA 19104}
\smallskip
\centerline{e-mail gould@payne.mps.ohio-state.edu}
\centerline{e-mail jordi@llull.physics.upenn.edu}
\bigskip
\centerline{\bf Abstract}
\singlespace 

  We propose that relative variability on short time-scales of the
multiple images of a lensed quasar, after removal of
the time delay, may be caused by hot spots or other moving structures
in the accretion disk crossing microlens caustics caused by
stellar mass objects in the lensing galaxy. Such variability has been
reported in the two images of 0957+561. The short durations would
be due to the high rotation speed of the disk $(v/c\sim 0.1)$, rather
than planetary mass objects in the slowly moving $(v/c\sim 10^{-3})$ lens.
This interpretation could be confirmed by finding periodicity, or
correlations of the spectral and flux variations due to the Doppler
effect in the disk.
We also propose another signature of stationary
accretion disks (with no intrinsic variability):
the gradient of the magnification over the accretion disk
should cause a relative color change between the images whose sign and
amplitude are correlated with the time derivative of the flux difference
between the images.
Other color terms
induced by the radial variation of disk colors are of second order in the
magnification gradient.
The methods proposed here can be used first to verify that accretion disks
near supermassive black holes are the source of the continuum radiation
from quasars, and then to study them.

\bigskip
\noindent Subject Headings: gravitational lensing -- quasars
\smallskip
\noindent Submitted to {\it The Astrophysical Journal Letters} December 13, 
1996
\smallskip
\noindent Preprint: OSU-TA-33/96
\endpage
\chapter{Introduction}

  For the past 30 years, the dominant view has been that the bulk of the
luminosity of quasars originates in
accretion disks spiraling in towards super-massive black holes
(Lynden-Bell 1969; see Rees 1984, Blandford et al.\ 1990,
Lin \& Papaloizou 1996 for reviews).
It is therefore striking that to date, there has been no clear observational
proof of this hypothesis.
The main features of the theory are:
1) The mass of the black hole is roughly estimated from its Eddington
luminosity: $M\gsim 10^8 M_\odot (L/10^{46} L_\odot)$.
2) The temperature of the accretion disk is estimated from the ``blue bump''
in the continuum spectrum to be $T\sim 5\times 10^4 $K.
3) The characteristic radius of the accretion disk is estimated assuming a
roughly black body intensity and is $\sim 10^3\,$AU for a bright quasar.
The main observational difficulty is that the angular size of the accretion
disk at a cosmological distance is then very small ($\sim 1\, \muas$),
and so has not been possible to resolve to date.

  Microlensing provides a potentially powerful probe of quasars since
the Einstein ring of a stellar-mass lens is typically $\sim 10^3\,$AU, i.e.\
of the same order as the expected size of the accretion disk. Microlensing
should occur in most lensed quasars. When multiple images of a quasar
are produced by a lensing galaxy (with typical separations of $1''$),
the total surface density of the lens must be
near the critical surface density. If a large fraction of this surface
density is contributed by stars, then these stars will no longer cause
microlensing lightcurves as isolated point masses, but instead their
caustic curves are linked in a complicated network. 
The lightcurve of a source moving behind a lens then results from
the variation of the magnification of several ``microimages'' with
separations of the order of the Einstein radii of the stars,
$\sim 1\, \muas$ (e.g., Wambsganss, Schneider, \& Paczy\'nski 1990),
and is therefore very complex.
Every time the source moves inside (outside) a caustic two new images
appear (disappear), and the magnification reaches a maximum that depends
on the size of the emitting region in the source.

  Microlensing was first detected by Irwin et al.\ (1989) in Huchra's Lens
(QSO 2237+0305; Huchra \etal 1985), and shortly thereafter by
Schild \& Smith (1990) in the original double quasar, 0957+561 (Walsh
\etal 1979). To verify that photometric
variation is due to microlensing, it must be distinguished from other
sources of variation, intrinsic variation in particular.
If one of the macroimages
is found to vary {\it relative} to the other, then one may conclude that one
or the other image (or both) are being microlensed.  However, the two images
arrive at the observer with a relative time delay.  Thus, the flux ratio
of the two images can be measured only by comparing observations that are
separated by this delay and, of course, this can only be done after the
time delay has been measured.

  If the timescale of a microlensing event is long 
compared to the time delay, the latter can be ignored to first approximation.
This is the case for 2237+0305: the time delays among the four images
are known to be of order days from the model of the lens, even though
they have not been measured. Since the microlensing events are of order
months, the time delay is not important.  Schild \& Choflin (1986) first 
measured the time delay for 0957+561 at $\tau=1.1\,$yr.  Schild \& Smith
(1990) used this delay when they showed that the B image had brightened 
relative to the A image by about 0.2 mag over 10 years.  This rise has 
subsequently slowed to a stop (R.\ Schild 1996, private communication).

  Intensive monitoring programs were then carried out to measure the
time-delay more precisely. 
Since the quasar varies on scales of one or two days at the few percent
level, a precision measurement did not at first appear difficult.
However, it was found by Schild
(1990) that there does not exist {\it any} value for $\tau$ that would
cause the 1-day to 3-month variations in the A and B images to coincide.
He and his collaborators
concluded that the underlying cause of most of these short timescale,
low amplitude ($\Delta m \sim 0.02$) 
variations is microlensing, not intrinsic variability (Schild 1990; 
Schild \& Smith 1991; Schild \& Thomson 1995; Schild 1996).
Kundi\'c \etal (1996) also reported finding such variations in the course of
measuring a more accurate time delay, $\tau=1.14 \pm 0.01\,$yr.

  The microlensing interpretation of the short timescale events proposed
by Schild has not received wide acceptance principally
because the short timescales would seem to imply that the dark matter
in the lensing galaxy is almost entirely composed of ``rogue planets''.
The argument for the mass scale is simple.
The characteristic physical Einstein radius is
$$r_E = \biggl({4 G M \dol\dls\over c^2 \dos}\biggr)^{1/2} ~, \eqn\rehat$$
where $\dol$, $\dos$, and $\dls$ are the angular diameter distances
between the observer, lens, and source. For $M=1 \msun$, and adopting
$H_0=55\kms\mpc^{-1}$ and $\Omega=1$,
$r_E/\dol = (6.2, 2.1) \muas$ for the lenses 2237+0305 and 0957+561,
respectively.
The crossing time is
$t_E = r_E/v_t$, where $v_t$ is the transverse velocity:
$$ \bvt = {\bvl - \bvo \over 1+z_l} - {\dol\over\dos}\, {\bvs - \bvo \over
1+z_s} ~. \eqn\vtrans $$
Here, $\bv_o$, $\bv_l$, and $\bv_s$ are the transverse velocities of the
observer, lens, and source, and $z_l$ and $z_s$ are the redshifts of the lens
and source
(Kayser, Refsdal, \& Stabell 1986; this formula is valid for flat
cosmological models, in general the observer's velocity relative to the
CMB frame must be multiplied by a factor that depends on the space
curvature, see Miralda-Escud\'e 1991). For 2237+0305 these parameters are 
$(z_l,z_s) = (0.039,1.69)$ and $(\dol,\dos,\dls) = (0.037,0.29,0.28)
c H_0^{-1}$, and for 0957+561 they are
$(z_l,z_s) = (0.36,1.41)$ and
$(\dol,\dos,\dls) = (0.21,0.30,0.18) c H_0^{-1}$.
The Sun's motion relative to the cosmic microwave background is 
$v_o\simeq 300\kms$, and if the transverse velocities of the lens and the 
source are similar then the
timescale to cross the Einstein radius for $M=1\msun$ is
$t_E\simeq (10$, $30)\,$yr for 2237+0305 and 0957+561, respectively.
Thus, when Irwin \etal (1989) reported a microlensing event with a
duration of 3 months in 2237+0305, the naive interpretation was that
this had been caused by a planet with a Jupiter mass.  For events with 
durations of a few days in 0957+561,
Earth masses would be inferred for the lensing objects.
Wambsganss et al.\ (1990)
pointed out that when the microlensing optical depth is large,
events with timescales much smaller than $t_E$ are possible (caused,
for example, by the passage of the source near a cusp catastrophe),
so low-mass lenses are not automatically implied 
from the observation of just
one event. However, the average density of caustic
crossings should still not be larger than $\sim$ 1 per Einstein radius,
so events on short timescales should be very rare unless planetary mass
objects account for a large fraction of the lens surface density.

  However, an additional difficulty with the planetary-mass 
interpretation of the short events
is the small size implied for the source. With the expected transverse
speeds given above, the quasar in 2237+0305 moves by only $0.2\,\muas$
relative to the lensing galaxy over three months, corresponding to
$\sim 300$ AU at the redshift of the quasar. This small size already
implies a rather high temperature for the disk emitting the continuum
($T\simeq 10^5$ K; Rauch \& Blandford 1991, see also Jaroszynski,
Wambsganss, \& Paczy\'nski 1992). For the shorter events in
0957+561, and with an expected angular velocity $\sim 10$ times smaller
than in 2237+0305 (due to the larger distance to the lens), the inferred
size of the source would be highly implausible within the framework of 
current models.

  In this paper, we propose that microlensing can be used to resolve
the structure of accretion disks. In \S\ 2 we examine the possibility
that short timescale events, of the type reported by Schild (1996), could be
produced by intrinsic variability in the accretion disk combined with
the effects of microlensing. In \S\ 3 we
turn to the case of a disk with no intrinsic variability, and propose a
method to demonstrate the existence of an accretion disk from the
spectral changes in the different macroimages as microlensing events
take place.

\chapter{Short Timescale Variability}

  The main difficulty in explaining the short timescale events is the
small proper motion expected, corresponding to the typical galaxy
velocities generated by large-scale structure of $\sim 300\, \kms$.
However, the orbital velocity of the accretion disk in the continuum
emitting region should be much larger. For example, if the black hole
mass is $M=10^8 \msun$, the circular velocity is $3\times 10^4 \kms$ at
a radius of $100$ AU, implying that the proper motion of any orbiting
blob is $(5, 40)$ times larger than the proper motion of the lensing
galaxy
in 2237+0305 and 0957+561 [from eq.\ \vtrans]. We therefore propose
that the short events observed in 0957+561 (and possibly some of the
variations in 2237+0305 as well) are caused by ``spots'' moving
at the typical orbital velocities. Such spots might be caused
by a variety of phenomena, including instabilities in the accretion
disk that create ``hot spots'' which are in roughly circular orbits, gas
clouds or stars that fall to and crash against the accretion disk,
or blobs ejected by a jet near the black hole.  They could even be ``cold
spots'', relatively confined subluminous regions like sunspots on the Sun.

  If the magnification of each image were uniform over the whole
accretion disk, then the intrinsic variability of the quasar caused
by such spots would be the same in all the images after
correction for the time delay, and therefore they would not be assigned
to microlensing. But if the magnification varies over the region of the
motion of a hot spot by $\Delta A$ due to microlensing, and the hot
spot contributes a fraction $\Delta F$ to the total quasar flux $F$,
then we expect a relative variation
of $(\Delta A/A)(\Delta F/F)$ of the flux in different images.
The timescale of these short events should be 
shorter than or
of order the dynamical time
in the accretion disk; for a radius $r=100$ AU and velocity
$v/c\sim 0.1$, this is about one month (which should be redshifted by
a factor $\sim 2.5$ for the two quasars we have mentioned).
If the spot is a transient phenomenon (changing its flux $\Delta F$ over
a dynamical time), 
we should expect an intrinsic variation of $\Delta F / F$, so the
intrinsic and relative variation 
should be correlated.

  The largest variations in magnification will occur when a caustic transits
the accretion disk. Two images of the accretion disk will then be
merging on a critical line and will be highly magnified. Let us assume
that the spot has a surface brightness higher by a factor $(1+f_s)$
compared to the average of the accretion disk ($f_s > 0$ for hot spots
and $f_s < 0$ for cold spots).
Then the radius of the
spot is $R_s = \left(\Delta F/(f_s F) \right)^{1/2} r$, where $r$ is
the radius of the disk. Since the maximum magnification on a fold
catastrophe is proportional to $R_s^{-1/2}$ (e.g., Schneider, Ehlers, \&
Falco 1992), the fractional contribution of a spot to the total flux
can rise to $(\Delta F / F)(r/R_s)^{1/2} = (|\Delta F| / F)^{3/4}\,
|f_s|^{1/4}$.  For example, a spot from a region contributing only 0.2\%
of the total light and with $|f_s|\sim 1$ could cause fluctuations
$\sim 1\%$.

\FIG\one{A representative example of the lightcurve that may result from
an orbiting hot spot. The upper panel shows the trajectory of the spot
over one orbit. The thick line represents the caustic (which is moving
upwards relative to the disk) and the dashed
line is the line of nodes of the orbit. The magnification near the caustic
is assumed to be $A = 1 + (b/y)^{1/2}$, where $y$ is the distance to the
caustic, and we choose the constant $b=2 \muas$, equal to the Einstein
radius of a $1\msun$ star. The radius of the orbit is
$r=100$ AU, inclined at an angle $50^o$ to the line-of-sight, and the
line of nodes has an inclination of $40^o$ relative to the caustic. The
orbital velocity of the spot is assumed to be $v/c = 0.1$, and the
relative proper motion of the caustic and the disk is 40 times smaller
than that of the hot spot. The solid line
in the middle panel shows the magnification of the spot as a function of
time, for a size of the spot equal to 0.1 times the radius of the orbit
(this size determines the maximum magnification at each caustic
crossing). The magnification can be measured
from the observed flux variation relative to another, non-microlensed
image after correction for the time-delay. The dotted line is the
actual flux variation of the spot, which is affected by intrinsic
variation due to Doppler effects when the spot is assumed to emit a
constant flux in its rest-frame. The lower panel shows the radial
velocity of the spot.
}


  For a spot that is either falling to or being ejected from the
black hole, the caustic will be crossed once. But if a spot is in
circular orbit and can survive for several orbits, then the caustic
will be crossed repeatedly. An example of the lightcurve that may result
from this is illustrated in Figure \one . The middle panel shows the
lightcurve for a single hot spot in circular orbit.
We assume that the caustic
crossing the disk is on a sufficiently large scale, so that the
magnification over the disk can be approximated as $A=A_0 + K/x^{1/2}$.
There is a characteristic pattern of variations as the spot crosses the
caustic: an ``M-shaped'' event is produced at every orbit of the spot.
These events might therefore be confused with similar events when the
motion is linear and two folds of the same closed caustic are crossed.
The periodic events become shorter until they disappear (the
time-reversed lightcurve is of course equally likely if the motion of the
disk is reversed). Such a spot would be likely to
have a different spectrum than the average for the quasar (it might,
for example, be much closer to a blackbody). The spectrum
is redshifted and blueshifted as the spot moves along the orbit (the
radial velocity is shown in the lower panel for our example). There
should then be a correlation between the spectrum and the magnification.
In our example, the spot is always more blueshifted at the beginning of
each event. This blueshift effect may be detectable even in the continuum
spectrum of a spot, given the large velocities involved.
Notice that such changes of the spectrum due to microlensing
of spots can be separated from any other intrinsic changes of the
spectrum of the entire disk, if one focuses on {\it relative} variations
of the spectrum between the multiple images of the quasar after
correcting for the time-delay.

	Whether an orbiting spot can survive for several orbits depends on 
the process that generates it.  A topologically confined spot, such as a
cold spot due to a magnetic vortex, may persist for many orbital periods.
Long lived spots may also be generated by gravitationally confined
disturbances in the gaseous disk, and by the effects of compact objects
orbiting inside the disk (such as black holes or other massive stars
that may have fallen to the disk).
On the other hand, hydrodynamical disturbances would expand in their
sound crossing time and would then be sheared out into
arcs by differential rotation. For example, if the ratio of the sound
speed to the circular velocity is 0.002 (e.g.\ $T= 5\times 10^4$ K
and $v/c=0.1$), a spot with orbital radius $r$
should, after $N$ periods, have expanded radially to a size
$0.01Nr$ due to expansion, and tangentially to a size $0.1 N^2 r$ due to
differential rotation. Therefore, such a spot cannot survive for more
than a few periods.

	In brief, spots can give rise to a variety of short timescale
signals in the difference between the flux from two quasar images.  The
deviations in the microlensed image minus the non-microlensed images can
be both positive and negative and can contain repeating, non-repeating, and
quasi-repeating signals.  

\chapter{Long Timescale Variability from a Steady Disk}

	The color of a microlensed accretion disk will in general be 
different from that of the disk in the absence of microlensing, due to
differential magnification over the disk.  If there are
two or more macroimages of the disk, then this color shift can be determined
by comparing the colors of the two images observed at two times separated
by the measured time delay.  The effect of microlensing can then be 
unambiguously distinguished from intrinsic color changes.  Color differences
between the images will arise if either or both are microlensed but for 
simplicity of discussion we will assume that one image is microlensed and the 
other is not.  

	Color changes are caused by two distinct effects of different orders.
Both require that the magnification vary over the source.  The variation can
be either smooth or due to a caustic.  We focus here on the case of smooth 
variations in the magnification because these are present generically.

	The first order effect arises from a coupling of the linear color 
gradient across the accretion disk (due to the disk's rotation) with the
gradient of the magnification.  This effect is analogous to the line shift
induced by microlensing of stars which has already been studied in some
detail (Maoz \& Gould 1994).  In the stellar case, the stellar lines are 
broadened by rotation because one side of the star is rotating toward the
observer and is blueshifted while the other side is redshifted.  If, for 
example, the blueshifted side of the star is more highly magnified, the
blue wing of the line will also be more highly magnified and the line 
centroid will shift toward the blue.  For a star with 
projected rotation speed $v\sin i$ and radius $r$, the shift is
$$\Delta v = {\zeta\over 4}r |\nabla \ln A|v\sin i\sin\gamma,\eqn\deltav$$
where 
$\nabla \ln A$ is the logarithmic gradient of the magnification and $\gamma$ 
is the angle between the 
magnification gradient and the projected axis of rotation.  The factor $\zeta$
depends on the details of the geometry.  For a line emitted
uniformly over the surface of a star, $\zeta=1$, while for a ring of emission
such as the Ca II line at 393 nm in giant stars (Loeb \& Sasselov 1995),
$\zeta=2$.

  The origin of the broad lines seen in virtually all quasars is a 
subject of considerable debate.  From their width ($v\sim 10^4\,\kms$),
they must come from regions that are at least
$\sim G M/v^2\sim 10^3 \,$AU from the center.
If they come from the accretion disk (or from any other structure with
organized motion on similar scales) then they will be subject to a line shift
given by equation \deltav.  While the exact value of the shift will depend
on the details of the geometry and will in addition be time dependent, the
general order of the effect should be $\sim v/4\sim 2000\,\kms$, assuming
$r |\nabla \ln A|\sim 1$. Notice that for $r=10^3$ AU, the angular size is
$\sim 1 \muas$; since the typical Einstein radius is only 1 to 5 times
larger, a magnification gradient $|\nabla \ln A|\sim 1/r$ should be common.
Hence, there should be an observable shift at some
times if the broad line region is indeed associated with the accretion disk.
Schild \& Smith (1991) measured the MgII line of the two images 0957+561
in two observations separated approximately by the time delay.
No difference in the lines was detected although it is not clear that
the data were of sufficient quality to see the predicted line shift.
In any event, a definitive test of the accretion-disk origin of the
broad lines would require measurements at multiple epochs.

	Even if the accretion disk does not give rise to broad lines, a 
magnification gradient can still generate a color shift.  In this case, the
entire quasi-thermal spectrum (``blue bump'') may be loosely considered as
a giant ``emission line'' with a fractional width of order unity.  More 
concretely, consider a color formed from the ratio of fluxes at two broad-band
wavelengths $\lambda_B$ and $\lambda_R$ with spectral slopes $\alpha_B$ and
$\alpha_R$.  By a calculation analogous to the one of Maoz \& Gould (1994),
the first order color shift $\Delta (B-R)_1$ is given by 
$$\Delta (B-R)_1 = {2.5\over \ln 10}\,|{\nabla \ln A}|\,
{\sin i\sin\gamma\over 2}\,
{\VEV{r v (\alpha_B -\alpha_R)}\over c},\eqn\deltabr$$
where $\VEV{r v (\alpha_B -\alpha_R)}$ is the intensity weighted mean over the
disk profile, and $\gamma$ is the angle between the minor axis of the projected
disk and the gradient of the magnification.  Thus, for 
$|\alpha_B-\alpha_R|\sim 1$, $\nabla \ln A\sim 10^{-3}\,\rm AU^{-1}$, and 
characteristic disk radius $\sim 300\,$AU, one might typically expect color 
shifts of order $\sim 1\%$.

  The second effect that causes color variations in microlensing on a
moving stationary disk, which is of second order, is due to the radial
color variation. This was investigated numerically by Wambsganss \&
Paczy\'nski (1991). The color change can be written, to second order,
as (see, e.g., Gould \& Welch 1996)
$$\Delta (B-R)_2\sim {\Lambda_B-\Lambda_R\over 8}(r\nabla\ln A)^2
,\eqn\deltabrtwo$$
where $r$ is the characteristic radius of the disk, and
$$ \Lambda_B = { 2\, \int_0^\infty dr'\, r'^3\, S_B(r') \over
r^2 \,\int_0^\infty dr'\, r'\, S_B(r')} ~, \eqn\biglam $$
is the second radial moment of the disk in the $B$ band (with the same
definition for the $R$ band).
If, for example, $\Lambda_B-\Lambda_R \simeq 0.1$, and
$(r\nabla\ln A) \sim 0.3$ (which we would expect typically),
this color term is still substantially smaller than the first order
term given by equation \deltabr.
The first order term, which is proportional to $\nabla \ln A$, then
dominates the relative color variations of the quasar images.
The time derivative of the magnification of the microlensed image is also 
proportional to $\nabla\ln A$: 
$${d\ln A\over d t} = v_t|\nabla \ln A|
\,{\dos\over \dol}\,{\cos\phi\over 1+z_l},
\eqn\timeder$$
where $\phi$ is the angle between the magnification gradient and the
transverse velocity.
Hence the first order color term should be directly proportional 
to the time derivative of the magnification ratio (after taking out the time 
delay).  This allows it to be distinguished from the second order term.
In practice, the proportionality between the color and the magnification
variation should not be perfect because, as the magnification varies, the
angle $\phi$ should also vary. Moreover, some of the color variations
may come from a sufficiently large region in the disk so that the
first-order term does not dominate. The maximum scale at which a
first-order expansion of the magnification can be applied is reduced, of
course, if low-mass stars or brown dwarfs are highly abundant, and is also
highly variable in the magnification patterns generated by the random
superposition of stars. Nevertheless, a correlation between the color
and magnification variation should clearly be present. Detailed
numerical simulations of lightcurves, with realistic disk models around
Kerr black holes (e.g., Jaroszynski \etal 1992) should help determine
the signatures of accretion disks that can be found.

{\bf Acknowledgements}: 
Work by AG was supported in part by a grant from the NSF 
AST 94-20746.

\endpage
\Ref\bland{Blandford, R. D., Netzer, H., \& Woltjer, L. 1990,
{\rm Active Galactic Nuclei} (Springer: Berlin) }
\Ref\mg{Gould, A.\ \& Welch D.\ L.\ 1995, ApJ, 464, 212}
\Ref\huc{Huchra, J. Gorenstein, M., Kent, S., Shapiro, I., Smith, G.,
Horine, I., \& Perley, R. 1985, AJ, 90, 691}
\Ref\irw{Irwin, M.\ J., Webster, R.\ L., Hewett, P.\ C.,
Corrigan, R.\ T., \& Jedrzejewski, R.\ I.\ 1989, AJ, 98, 1989}
\Ref\mg{Jaroszynski, M., Wambsganss, J., \& Paczy\'nski, B. 1992, ApJ, 396, 
L65}
\Ref\kay{Kayser, R., Refsdal, S., \& Stabell, R. 1986, A\&A, 166, 36}
\Ref\kun{Kundi\'c, T., et al.\ 1996, submitted to ApJ (astro-ph 9610162)}
\Ref\mg{Lin, D. N. C., \& Papaloizou, J. C. B. 1996, ARA\&A, 34, 703}
\Ref\lyn{Lynden-Bell, D. 1969, Nature, 223, 690}
\Ref\mg{Loeb, A.\ \& Sasselov, D.\ 1995, ApJ, 449, L33}
\Ref\mg{Maoz, D.\ \& Gould, A.\ 1994, ApJ, 425, L67}
\Ref\mir{Miralda-Escud\'e, J. 1991, ApJ, 379, 94}
\Ref\rau{Rauch, K. P., \& Blandford, R. D. 1991, ApJ, 381, L39}
\Ref\rees{Rees, M. J. 1984, ARA\&A, 22, 471}
\Ref\ss{Schild, R.\ E.\ 1990, AJ, 100, 1771}
\Ref\ss{Schild, R.\ E.\ 1996, ApJ, 464, 125}
\Ref\ss{Schild, R.\ E.\ \& Choflin, B.\ 1986, AJ, 98, 1989}
\Ref\ss{Schild, R.\ E.\ \& Smith, R.\ C.\ 1991, AJ, 101, 813}
\Ref\ss{Schild, R.\ E.\ \& Thomson, D.\ J.\ 1995, AJ, 109, 1970}
\Ref\schn{Schneider, P., Ehlers, J. \& Falco, E.~E. 1992, {\rm
Gravitational Lenses} (Berlin: Springer-Verlag)}
\Ref\ss{Walsh, D., Carswell, R. F., \& Weymann, R. J. 1979, Nature, 279, 381}
\Ref\ss{Wambsganss, J., \& Paczy\'nski, B. 1991, AJ, 102, 864}
\Ref\ss{Wambsganss, J., Schneider, P., \& Paczy\'nski, B. 1990, ApJ, 358, L33}

\refout
\endpage
\figout
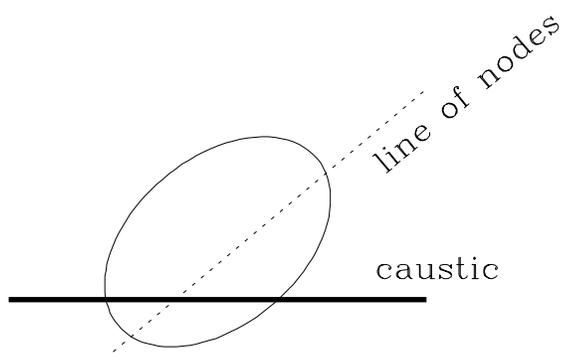
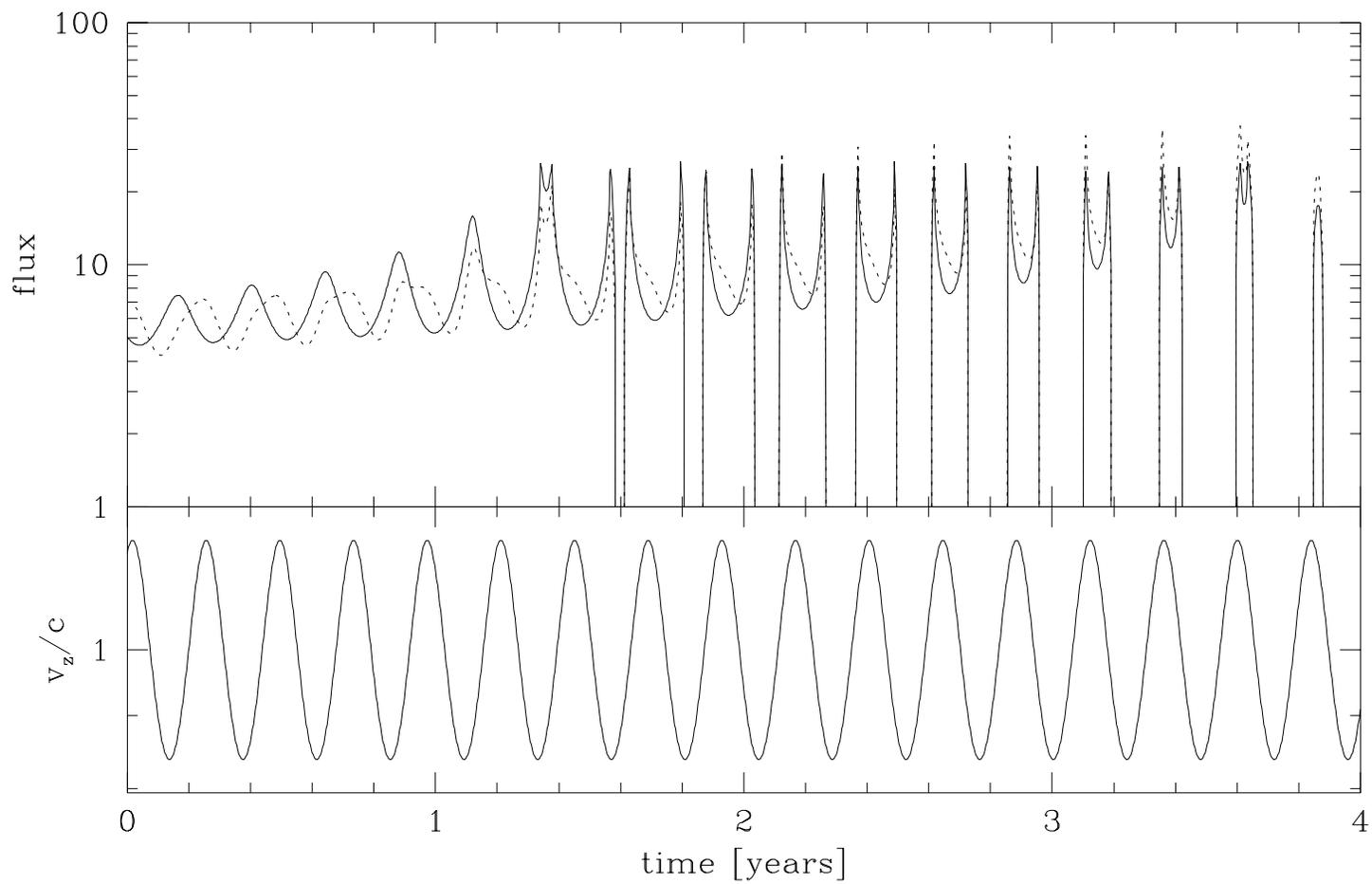
\end